%% file: pamTrust.tex
\documentclass[letterpaper, 10 pt, conference]{ieeeconf}  

\IEEEoverridecommandlockouts                              
\usepackage{tablefootnote}
\usepackage{amssymb}
\usepackage{amsmath}
\usepackage{graphics}
\usepackage{graphicx}
\usepackage{epsfig}
\usepackage{epstopdf}
\usepackage{mathrsfs}
\usepackage{cite}
\usepackage{subcaption}
\usepackage[mathscr]{euscript}


\newtheorem{theorem}{Theorem}[section]

\newtheorem{corollary}[theorem]{Corollary}

\begin{document}
\title{The Philos Trust Algorithm: Preventing Exploitation of Distributed Trust}
\author{{Pam Russell and Philip N. Brown}
\thanks{P. Russell and P. N. Brown are with the Department of Computer Science at the University of Colorado Colorado Springs (UCCS) {\texttt{\{prussell,philip.brown\}@uccs.edu}. This work is supported in part by Colorado State Bill 18-086.}}
}

\maketitle              
\begin{abstract} 

The Philos Marketplace blockchain system is a proposed hierarchical blockchain architecture which allows a large number of individual blockchains to operate in parallel.
These parallel chains achieve consensus among one another on a limited set of core operations, while allowing each on-chain application to manage its own data independently of others.
This architecture addresses the scalability issues of traditional linear blockchains, but requires novel consensus mechanisms.
A central feature of the Philos consensus mechanism is its \emph{trust algorithm,} which assigns each network node a numerical trust value (or score) indicating the quality of recent past performance.
This trust value is then used to determine a node's voting weight at the higher levels of consensus.
In this paper, we formally define the Philos trust algorithm, and provide several illustrations of its operation, both theoretically and empirically.
We also ask whether a misbehaving node can strategically exploit the algorithm for its personal gain, and show that this type of exploitation can be universally prevented simply by enforcing a mild limit on the number of participants in each of the parallel chains.

\end{abstract}

\section{Introduction}

A blockchain is a decentralized, shared, and immutable digital ledger that records transactions in blocks that are linked together cryptographically.
Beginning with Bitcoin, purpose-built blockchains have been proposed for a wide variety of applications including payment networks~\cite{Nakamoto2008,Ben-Sasson2014}, decentralized provision of DNS information~\cite{Chang2016,liu_data_2018}, gambling~\cite{Du2019}, social media~\cite{Li2019}, stablecoins~\cite{Brown2019e,Klages-Mundt2020} and many others~\cite{Lasla2018}.
This early array of purpose-built blockchains was followed by a wave of blockchain projects which aim to be general-purpose in nature, allowing features like Turing-complete on-chain scripting and other highly customizable functionality~\cite{Tikhomirov2018,Li2020,Androulaki2018,Zheng2021}.

These blockchain systems operate by defining a {consensus mechanism}, a set of rules defining how the various participants agree on which transactions and operations are valid and should be included in the permanent ledger.
These range from the computationally intensive proof-of-work systems pioneered by Bitcoin~\cite{Nakamoto2008}, to newer systems such as proof-of-stake and its many variants~\cite{King2012,Bentov2014,Kiayias2017}.
Central to any consensus mechanism must be a guarantee that blockchain participants are properly incentivized to participate according to the nominal design of the mechanism.
Unfortunately, this type of incentive compatibility can be challenging to prove, and is known to be weak on Bitcoin itself as evidenced by the many known mining difficulty exploits~\cite{Eyal2015,Sapirshtein2017,Kwon2017,Elliott2019,Negy2020}

Scalability is another significant challenge which still faces blockchain technology.
In many currently-operating mainstream blockchain implementations, the entire history of all on-chain activity is replicated across all participating nodes.
This leads to immense databases for very old (such as Ethereum, $\sim$1 TB) or very active blockchains (such as EOS, which grew to $\sim$4 TB in its first 4 months of operation).
The blockchain scalability challenge has led to many proposed mitigation systems which use sidechains~\cite{agarwal_blockchain_2020}, database sharding~\cite{Chauhan2018,Dang2019}, and other techniques to limit the amount of data which must be replicated across the entire network.

One such proposed system is the Philos Marketplace blockchain system~\cite{gorog2018},\cite{gorog2021patent}.
Philos attempts to mitigate the scalability problem by employing a hierarchical consensus model which allows a large number of individual blockchains to operate in parallel.
Each of these individual chains is envisioned to support a single application, and to be maintained by a relatively small number of known network nodes which collectively are called a \emph{sync list} through a strict synchronous consensus process. We will refer to these nodes as \emph{peer}.

Higher in the hierarchy, these parallel chains achieve consensus among one another on a limited set of core operations, while allowing each on-chain application to manage its own data independently of others.
Thus, the data maintained by a particular parallel chain need not be completely replicated across the entire network of parallel chains, while cross-chain transactions can still be executed and validated at a global network level. 
This architecture addresses the scalability issues of traditional linear blockchains, but in doing so it requires novel consensus mechanisms.

To address this need, this paper presents the Philos Trust Algorithm, a decentralized means of assigning a \emph{trust value} or reliability score to each participating peer; a peer's trust value is then used to determine its weight in the on-chain consensus mechanisms. 
In this paper, we provide the first formal description of the Philos Trust Algorithm, provide several examples to illustrate its functioning, and analytically prove several important properties that it possesses.
First, we show that there is a fundamental upper bound on the trust value which a peer can only achieve by following the protocol exactly. 
That is, if a peer deviates from the protocol in any way, this necessarily results in reductions in the trust value relative to the theoretical maximum.
Second, we show that a particular hypothetical exploit on the Philos Trust Algorithm can be universally prevented simply by enforcing a simple condition on the size of sync lists.

The goal of this paper is twofold: 
First, in Section~\ref{sec:trust algo}, we explicitly define the algorithm which maintains and updates this trust value, and provide several illustrative examples of how this works.
Second, in Section~\ref{sec:incent}, we investigate the incentive effects of trust.
Specifically, we show how to prevent a specific strategic exploit of the trust algorithm simply by limiting the maximum size of sync lists.
For context, Section~\ref{sec:philos} begins with a high-level overview of the basic operation of the Philos system; Section~\ref{sec:trust algo} presents the Philos Trust Algorithm in detail and gives the examples and upper bound result, and Section~\ref{sec:incent} gives the incentive compatibility results.

\section{Philos Marketplace Blockchain System}\label{sec:philos}
A blockchain is a shared database that stores data in blocks and links them together using cryptography. Transactions are recorded chronologically, forming an immutable chain. The chronological adding of transactions to the blockchain has raised questions about its future scalability and energy consumption. Several key metrics have been analyzed to measure the scalability of Bitcoin with two of the most important performance metrics being maximum throughput and latency. Since transaction throughput is restrained by block interval and the block size, a larger block can store more data but it increases the propagation time. Transaction latency is the time for a transaction to be confirmed. The more transactions, the longer it takes for transactions to be accepted~\cite{zhou_solutions_2020}.

The Philos Marketplace Blockchain is designed with scalability in mind and does so by allowing an expandable system of parallel blockchains.
These blockchains are loosely managed by a small group of trusted consortium servers and a complex consensus mechanism. 
Basic consistency among these parallel chains is enforced by occasional global consensus operations known as \emph{bridge syncs}.
However, most of the transaction activity occurs at a lower level of consensus on the parallel chains themselves, whose local consensus is enforced by regular local synchronization events called \emph{primary syncs}.
This addresses both throughput and latency concerns, since these local primary syncs are only occasionally propagated globally in bridge syncs, and then only as block hashes.
Thus, scalability is improved by reducing the number of transactions being replicated across all peers.  

The basic server type that runs the Philos Marketplace blockchain is called a \emph{peer.} To interact with the blockchain system, a peer first groups together with two or more other peers to form a \emph{sync list}.
At regularly scheduled intervals (typically every 10 minutes), all peers in the sync list collaborate to create and cryptographically sign a \emph{primary block,} which provisionally locks in the previous 10 minutes of the sync list's activity.
To incorporate these provisional primary blocks into global consensus, the sync list must occasionally participate in a consensus operation with other sync lists; this higher-level operation is known as a \emph{bridge sync} (typically every 8 hours).
When a sync list participates in a bridge sync, every peer in every participating sync list agrees to permanently commit all provisional primary blocks to the global blockchain.
Hence, the bridge sync is a critical operation. 

\begin{figure}
    \centering
    \includegraphics[width=8cm]{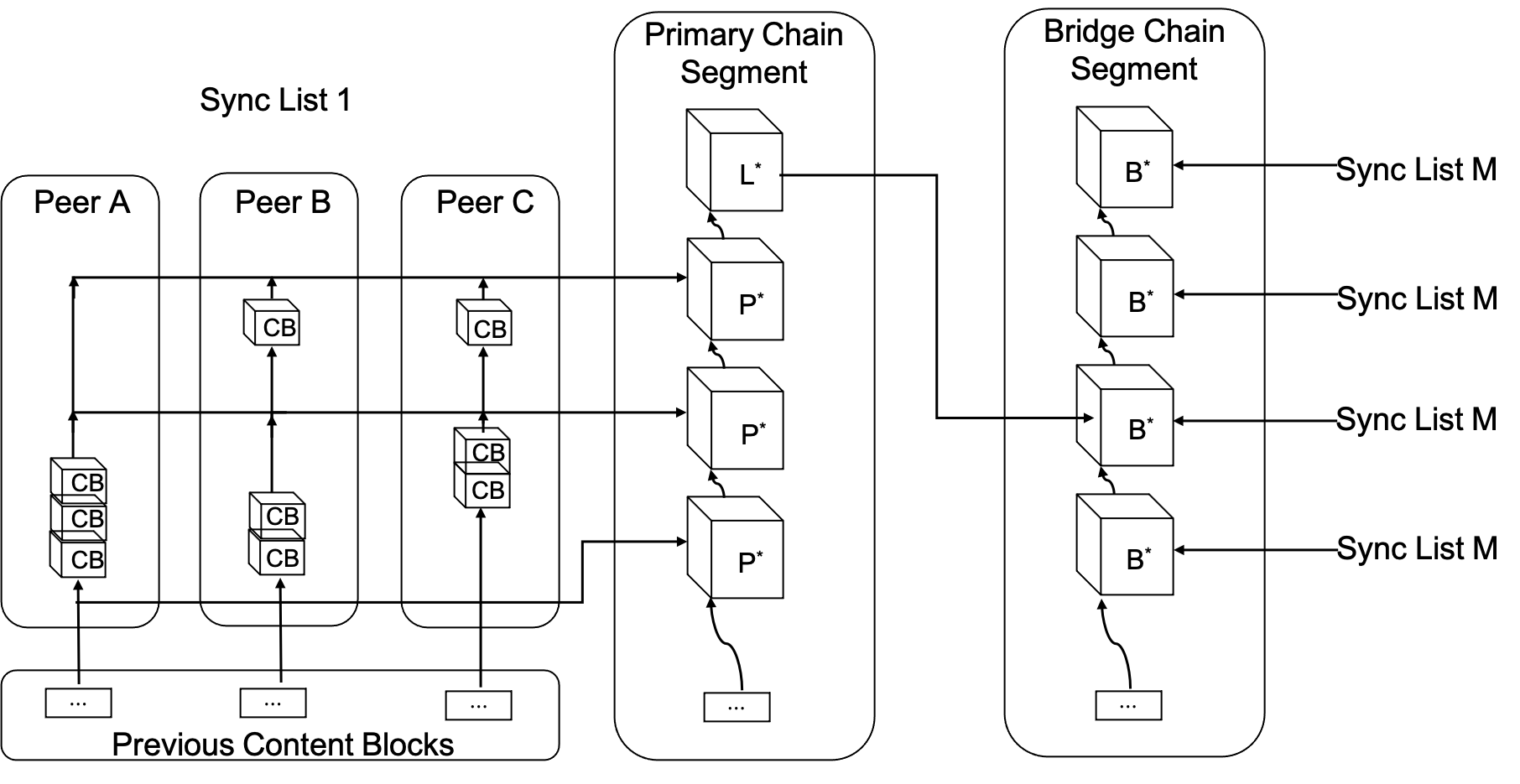}
    \caption{\footnotesize \label{fig:PrimaryProcess}Primary Consensus Process: Peers in a sync list create parallel content blocks $CB$. At specific periods called the prime step (typically 10 minutes), these content blocks are aggregated into a primary consensus block $P^*$ in a process called a primary sync. Sync lists continue to perform primary syncs until they decide to join a bridge consensus with other sync lists by creating a link block $L^*$ and then join a bridge consensus block $B^*$ which is broadcast globally. Syncs lists can then resume activity or request a dissolution of the sync list.}
\end{figure}

Illustrated in Figure~\ref{fig:PrimaryProcess} is the primary consensus process performed by the sync list to participate in a bridge sync. Each peer maintains a parallel blockchain consisting of content blocks CB which are used to create the primary block $P^*$. This process is called a \emph{primary sync}. The primary sync process continues until the sync list wishes to participate in the bridge sync. They do so by creating a link block $L^*$ which aggregates the relevant information in conjunction with the $P^*$ blocks and then they will join a bridge consensus with other sync lists, thus creating a bridge block $B^*$ which is broadcast globally. At the completion of the bridge sync, sync lists can choose to resume activity and continue to make content blocks or they can request a dissolution of the sync list to dissolve. 

While the full Philos system is considerably more complex and contains many features which are out of scope of this paper, this brief overview provides enough context to make the paper's contributions clear.


\subsection{Assessing Peer Trustworthiness}

The Philos Marketplace blockchain employs a trust-based global consensus model to incentivize peers to participate faithfully in consensus among their individual sync lists.
This system calculates a running \emph{trust} value (or score) for each peer to measure reliability and responsibility at the primary consensus level.
In turn, a peers ability to participate in global consensus (e.g., bridge sync) is weighted by its trust value.
That is, a bridge block is only considered valid if it is signed by peers which, put together, control a minimum fraction of the sum of all peer trust values in the system.
This ensures that blocks entering global consensus have been vetted and approved by sufficiently well-regarded members of the community.
Accordingly, each peer is incentivized to maintain a high trust value (that is, participate faithfully in primary consensus) so that it can maximize its weight (essentially voting power) in the bridge consensus step.

\section{Trust Value Algorithm}\label{sec:trust algo}

\subsection{Basics}

The purpose of the trust value algorithm is to provide a mathematical means of calculating a trust value for each peer indicating the level of certainty that a peer will reliably participate in consensus operations. Trust values will increase with successive successful consensus operations and will decrease when consensus operations are missed.


To reward peers with an increase in trust value, the number of primary sync opportunities a peer has acquired since their last bridge sync is kept and used in the trust value algorithm. As long as the sync list in which a peer is associated with performs according to the chains consensus operation, trust values will increase. Simultaneously, if a sync list misses a primary consensus or a bridge consensus, the trust value of each peer in the sync list will decrease.

\subsection{Algorithm}

The trust value of each peer is calculated and updated after inclusion of the primary block in a bridge sync.
The calculation is a recursive algorithm that incorporates a value for the current calculated trust as well as a decay on any previous trust. The idea is to reward peers for reliable participation in successful primary syncs. In addition, the trust value considers unsuccessful primary syncs by creating a hidden decay that is applied when the peer performs their next bridge consensus. This failure to complete a primary consensus can be either unintentional or deliberate. In either case, all peers within a sync list that fail to perform a primary consensus will realize a reduction in their trust value. However, over time, if an individual peer reliably performs successive primary syncs, this decay mechanism gradually reduces the penalty of old bad behavior. This allows for peers to be rewarded and ensures that past mistakes are not remembered forever and will decay over a period of time. 

Because the primary consensus is synchronous, we henceforth express all measures of time in units of Primary Consensus intervals (or \emph{prime steps}) which we index by $k$. This is a global counter indicating the number of primary consensus intervals that have elapsed since the creation of the chain.
We write $\Delta$ to denote the \emph{max bridge interval}. The max bridge interval represents the maximum number of primary syncs that can be performed by a sync list within the given bridge consensus interval. 
For ease of reference, we provide Table~\ref{table:notation} to assist the reader in following the mathematical notation.
We keep the mathematical description as general as possible, but provide a listing of reference parameter values in Section~\ref{ssec:ref values}.

\begin{table}[h]
    \centering
     \caption{Table of notation}
    \begin{tabular}{c p{0.8\columnwidth}}\label{table:notation}

       Variable  & Definition\\
       \hline
       $N$          & Set of all peers \\
       $M$          & Sync list; $M\subset N$\\
    $i$             & Peer index \\
    $b_k^i$         & Time stamp of peer $i$'s most recent bridge as of time $k$ \\
    $k$             & Block index (unit: prime step) \\
    $\Delta$        & Max bridge interval \\
    $S_M$         & Number of successful primaries by sync list $M$ since previous bridge\\
    $T_i$         & Peer $i$'s raw trust value at time $k$ \\
    $\bar{T}_i$   & Hypothetical trust a peer would have if trust updated with $S_M=0$ \\
    $\mathcal{T}_i$ & Peer $i$'s fractional trust value at time $k$ \\
    $k-b_k^i$       & Primary sync counter ($i$'s number of primary opportunities since it's last bridge)\\
    $\beta$         & Exponential decay constant
    \end{tabular}
    \label{tab:notations}
\end{table}

When sync list $M$ bridges between prime steps $k$ and $k+1$, the trust value $T_i$ is calculated for each peer ledger $i\in M$.
This calculation uses the peer's previous trust value $T_i^-$, the timestamp of the peer's most recent bridge $b_k^i$, and $S_M$, the number of successful primary syncs recently performed by the sync list.
Given these and an exponential decay constant $\beta$ which satisfies $0<\beta<1$, the new trust value  $T_i$ is computed according to the following:
\begin{equation} \label{eq:trustvalue}
T_i := T_i^-\beta^{k-b_k^i} + S_M\min\left\{1, \left(\frac{k-b_k^i}{\Delta}\right)\right\}.
\end{equation}

Note in~\eqref{eq:trustvalue} that the first term in the sum captures the decay of the old trust; a long time since the last bridge (i.e., a large value of $k-b_k^i$) causes a larger decay of trust.
The second term in the sum captures the reward associated with recent primary syncs by peer $i$'s sync list; larger values of $S_M$ garner larger rewards.
Finally, the $\min\{\cdot\}$ term penalizes sync lists which bridge much sooner than the max bridge deadline $\Delta$; this is because over-frequent bridging leads to blockchain bloat. 
It should be noted that when a peer first joins the system, its trust value is initialized to zero. 

The trust value aims to predict the reliability of a peer to successfully perform primary syncs. As primary syncs are the basic building block of the blockchain, the reliability of a peer will determine the successful matching of ledgers to a sync list and thereby, increase the probability that bridge syncs can be performed. 

We define the \emph{fractional trust} of peer $i$ as
\begin{equation}
    \mathcal{T}_i := \frac{T_i}{\sum_{\ell\in N}T_\ell},
\end{equation}
which represents the percentage of trust that peer $i$ has in relation to the total system trust. 
This is the quantity of highest interest to a peer, since bridge consensus requires a minimum threshold of the total system trust. Thus, it is reasonable to assume that all peers desire to have a high fractional trust value.

\subsection{State Machine to Model Consensus Process}

\begin{figure}
    \centering
    
    \includegraphics[width=0.45\textwidth]{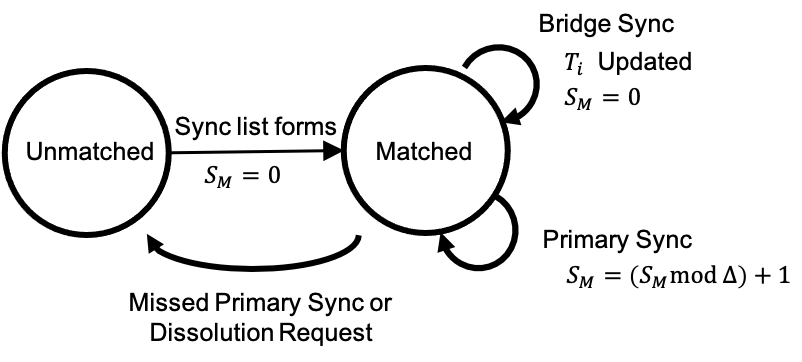}
    
    {\footnotesize
    \begin{tabular}{c|c|c}

    Transition & Cause & Effect\\
       \hline 
       Sync List Forms      & unanimous       & $S_M=0$ \\
                             & asynchronous  \\        
        \hline
        Primary Sync    & unanimous    & $S_M=(S_M \bmod{\Delta}) +1$ \\
                        & synchronous \\
        \hline 
       Bridge Sync  & unanimous             & Trust $T_i$ updated \\
                        & asynchronous      & according to equation (1) \\
        &   & $S_M=0$\\
        \hline
       Missed primary sync  & unilateral          & Sync list disbands \\
                        & synchronous  \\
       \end{tabular}}
    \caption{\footnotesize\label{fig:statemachine}Peer State Machine: Peers start off in an unmatched state and are matched to form a sync list, and the sync list primary counter $S_M$ is initialized to zero. After matched with a sync list, peers participate in a sequence of (synchronous) primary or (asynchronous) bridge syncs. If a sync list misses a scheduled primary sync either due to deliberate sabotage by a peer or an unintended fault, the sync list is automatically dissolved and the peers are returned to the unmatched state. If a sync list performs a primary sync, the sync list primary counter is incremented and the peers continue activity. If a sync list performs a bridge sync, the trust value for each peer is updated and the sync list primary counter is initialized to zero. Sync lists can also request dissolution and be returned to the unmatched state.}
\end{figure}

The process of sync list formation, primary sync, bridge sync, and sync list dissolution is depicted in the state transition diagram depicted in Figure~\ref{fig:statemachine}.
We describe each state transition in detail below.
For specific examples of timing and how these operations interact with trust values, see also Section~\ref{ssec:examples}.

\subsubsection{Sync List Forms}
This transition occurs when a set of unmatched peers unanimously agree to form a sync list $M$ together. Once a sync list is formed, the primary sync counter is initialized to zero: $S_M=0$. This counter accumulates the number of successful primary syncs performed between bridging and is used in the trust value calculation~\eqref{eq:trustvalue}. 

\subsubsection{Primary Sync}
Once peers have been matched and form a sync list $M$, the sync list begins to add data to its own parallel blockchain and performs a consensus amongst the group creating the primary consensus $(P^*)$ block known as a primary sync. The protocol requires that primary syncs occur at every prime step $k$. Each time a sync list performs a primary sync, the primary sync counter $S_M$ is incremented. However, sync lists that do not bridge at least every $\Delta$ prime steps are penalized by resetting $S_M$ to $1$ when the max bridge interval is exceeded.

\subsubsection{Bridge Sync}
To lock in the primary syncs performed, the sync list performs a bridge sync at any time between primary syncs. The trust value for each peer in the sync list is then calculated and updated according to~\eqref{eq:trustvalue}, crediting each peer based on $S_M$, and the primary sync counter is reset $S_M=0$. The sync list can continue to remain as a sync list and perform new primary syncs or they can choose to disband and return to the unmatched state. 

\subsubsection{Missed Primary Sync}
This transition occurs when an individual peer fails to participate in a primary sync, deliberately or not. The sync list is disbanded and the blockchain of each of the list's peers is reverted to the most recent state locked in by the previous bridge consensus. In contrast, if a sync list performs a successful bridge sync, they can submit for a dissolution request of the sync list and be returned to the unmatched state,

\subsection{Equilibrium Trust}

Given the state transition model described in Figure~\ref{fig:statemachine} and in the previous subsection, it is possible to calculate the equilibrium trust value that a peer would have if it performed perfectly for a very long time.
That is, the peer's sync list never misses a primary sync, and also consistently waits until the max bridge interval $\Delta$ to perform a bridge sync.
Furthermore, this equilibrium value is the largest trust value possible in the system.
This is formalized in the following theorem:

\begin{theorem} \label{theorem:equilibrium}
If a peer $i$ performs perfectly (i.e. $S_M=\Delta$ and $k-b_k^i=\Delta$) at every time step, then the following hold:
\begin{enumerate}
    \item Trust asymptotically approaches the equilibrium trust value of
        \begin{equation}  \label{eq:equiltrust}
        T^* = \frac{\Delta}{1-\beta^\Delta},
        \end{equation}
    \item Trust increases at every bridge: if $T_i^-<T^*$, then
        \begin{equation} \label{eq:trust increases}
        T_i > T_i^-.
        \end{equation}
\end{enumerate}
Furthermore, regardless of the behavior of peer $i$, trust never exceeds $T^*$:
    \begin{equation}\label{eq:tstar bound}
        T_i \leq T^*.
    \end{equation}
\end{theorem}\vspace{2mm}

\noindent The proof of Theorem~\ref{theorem:equilibrium} is presented in the Appendix.

\subsection{Calculating Reference Parameter Values} \label{ssec:ref values}
There are several parameters that are set by the chain and are envisioned to be a one-time occurrence. While modifications may need to be made at the start of the system, they are not envisioned as a value that would be constantly updated.
Table~\ref{table:parameters} provides a list of reasonable numerical values for the chain default parameters.
Here, $K$ denotes the length of a prime step; that is, the time interval between time indices $k$ and $k+1$.


\begin{table}[h]
    \centering
     \caption{Table of Reference Parameters}
    \begin{tabular}{c|c|c
    p{0.6\columnwidth}}\label{table:parameters}

       Parameter  & Reference Value & Description\\
       \hline
       $\beta$      & $0.9999111696$ & Decay base \\
       \hline
       $\Delta$          & $48$ & Bridge should \\
                        & & occur at most every 8 hours\\
       \hline
       $K$          & 10 minutes & Primary sync \\
                    & & occurs every 10 minutes\\
       \hline
       $m$        & 6 months & Number of months to \\
                & & reach $P$ fraction of equilibrium trust \\
       \hline
       $P$        & 90\%  & Percent of equilibrium \\
            & & trust to be reached \\
    \end{tabular}
    \label{tab:notations}
\end{table}

The decay base $\beta$ is calculated by choosing how many months it takes for a peer to obtain a certain percentage of the equilibrium trust value.
That is, if the chain designer wishes for it to take $m$ months for a ledger to reach a trust value of $PT^*$, the required $\beta$ is given by the formula

\begin{equation}
    \beta=(1-P)^{\frac{K}{m(60)(24)(30)}}.
\end{equation}

The particular $\beta$ in Table~\ref{table:parameters} was calculated by choosing that peers reach $P=90\%$ of the equilibrium trust in $m=6$ months. In the equation for $\beta$, the number of months \emph{m} is converted to the number of primaries it takes to reach the equilibrium trust by multiplying the number of months by 30 days in a month, 24 hours in a day, and 60 minutes in an hour.

For example, if it is desired that peers approach $90\%$ of the equilibrium trust more slowly, say in one year, $\beta$ would be calculated as follows:
\begin{align}
    \beta&=(1-0.9)^{\frac{10}{12(60)(24)(30)}},\\
    \beta&=0.9999555838.
\end{align}


%

\begin{figure}
    \centering
    \includegraphics[width=8cm]{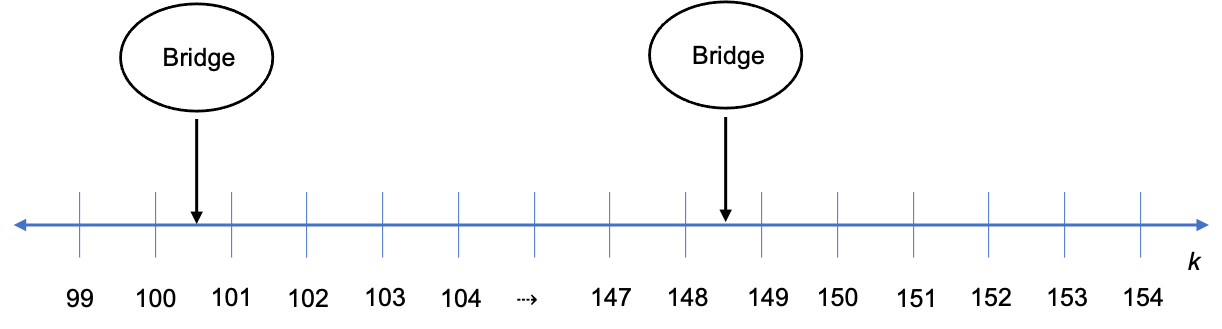}\vspace{2mm}
    
    {\footnotesize
    
    \begin{tabular}{c|c|c|c|c}

    Prime Step $k$ & $b_k^i$ & $k-b_k^i$ & $S_M$ & Trust Value Calc \\
       \hline 
       $98$ & $52$  & $46$       & $46$ \\
        \hline
       $99$ & $52$  & $47$       & $47$ \\ 
        \hline 
       $100$ & $52$  & $48$       & $48$\\
        \hline
         & & & & $T_i^-\beta^{48} + 48(1)$ \\
         \hline
       $101$ & $100$  & $1$       & $1$\\
        \hline
       $102$ & $100$  & $2$       & $2$\\
        \hline
       $103$ & $100$  & $3$       & $3$\\
        \hline
       $104$ & $100$  & $4$       & $4$\\
       \hline 
       \vdots & \vdots &\vdots &\vdots &\vdots  \\
        \hline
       $147$ & $100$  & $47$       & $47$\\
        \hline
       $148$  & $100$  & $48$       & $48$\\
        \hline
         & & & & $T_i^-\beta^{48} + 48(1)$ \\
         \hline
       $149$ & $148$ & $1$       & $1$ \\
        \hline
       $150$ & $148$  & $2$       & $2$ \\
        \hline
       $151$ & $148$  & $3$       & $3$ \\
        \hline
       $152$ & $148$ & $4$       & $4$ \\
        \hline
       \end{tabular}
    }
    \caption{\footnotesize\label{fig:normalexample}Perfect Bridging Example: The sync list bridges at the max bridge intervals (i.e. between prime steps 100 and 101 and prime steps 148 and 149).}
\end{figure}

\subsection{Illustrative Examples}\label{ssec:examples}
\begin{figure}
    \centering
    \includegraphics[width=8cm]{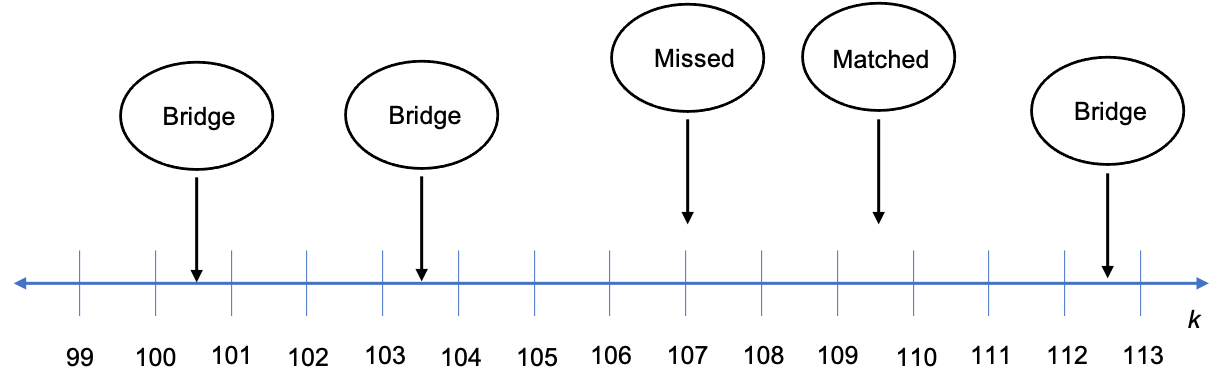}\vspace{2mm}
    
    {\footnotesize
        \begin{tabular}{c|c|c|c|c}
    
    Prime Step $k$ & $b_k^i$ & $k-b_k^i$ & $S_M$ & Trust Value Calc \\
       \hline 
       $98$ & $52$  & $46$       & $46$ \\
        \hline
       $99$ & $52$  & $47$       & $47$ \\ 
        \hline 
       $100$ & $52$  & $48$       & $48$\\
       \hline
         & & & & $T_i^-\beta^{48} + 48(1)$ \\
        \hline
       $101$ & $100$  & $1$       & $1$ \\
        \hline
       $102$ & $100$  & $2$       & $2$ \\
        \hline
       $103$ & $100$  & $3$       & $3$ \\
       \hline
         & & & & $T_i^-\beta^{3} + 3(\frac{3}{\Delta})$ \\
        \hline
       $104$ & $103$  & $1$       & $1$\\
        \hline
       $105$ & $103$  & $2$       & $2$\\
        \hline
       $106$ & $103$  & $3$       & $3$\\
        \hline
       $107$ & $107$  & $4$       & $--$\\
        \hline
       $108$ & $108$  & $5$       & $--$\\
        \hline
       $109$ & $109$  & $6$       & $--$\\
        \hline
       $110$ & $103$ & $7$       & $1$ \\
        \hline
       $111$ & $103$  & $8$       & $2$ \\
        \hline
       $112$ & $103$  & $9$       & $3$ \\
        \hline
         & & & & $T_i^-\beta^{9} + 3(\frac{9}{\Delta})$ \\
         \hline
       $113$ & $112$ & $1$       & $1$ \\
        \hline

       \end{tabular}}
    
    \caption{\footnotesize\label{fig:earlyexample}Early and Missed Bridging Example: The sync list bridges at the max bridge interval, between prime step 100 and 101, bridges again between prime step 103 and 104 (too early), misses a primary sync at prime step 107, is returned to unmatched state, is matched with new sync list between prime step 109 and 110, and bridges too early between prime step 112 and 113.}
\end{figure}

\begin{figure}
    \centering
    \includegraphics[width=8cm]{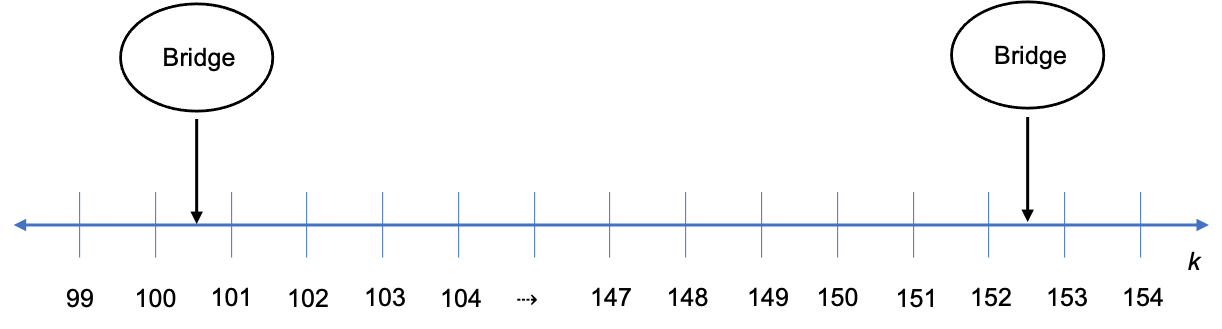}\vspace{2mm}
    
    {\footnotesize
        \begin{tabular}{c|c|c|c|c}
        Prime Step $k$ & $b_k^i$ & $k-b_k^i$ & $S_M$ & Trust Value Calc \\
       \hline 
       $98$ & $52$  & $46$       & $46$ \\
        \hline
       $99$ & $52$  & $47$       & $47$ \\ 
        \hline 
       $100$ & $52$  & $48$       & $48$\\
        \hline
         & & & & $T_i^-\beta^{48} + 48(1)$ \\
         \hline
       $101$ & $100$  & $1$       & $1$\\
        \hline
       $102$ & $100$  & $2$       & $2$\\
        \hline
       $103$ & $100$  & $3$       & $3$\\
        \hline
       $104$ & $100$  & $4$       & $4$\\
        \hline
       $147$ & $100$  & $47$       & $47$\\
        \hline
       $148$  & $100$  & $48$       & $0$\\
        \hline
       $149$ & $100$ & $49$       & $1$ \\
        \hline
       $150$ & $100$  & $50$       & $2$ \\
        \hline
       $151$ & $100$  & $51$       & $3$ \\
        \hline
       $152$ & $100$ & $52$       & $4$ \\
        \hline
         & & & & $T_i^-\beta^{52} + 4(1)$ \\ 
         \hline
       $153$ & $152$ & $1$       & $1$ \\
        \hline
       $154$ & $152$ & $1$       & $1$ \\
        \hline
       \end{tabular}
    }
    
    \caption{\footnotesize\label{fig:lateexample}Late Bridging Example: The sync list bridges at the max bridge interval, between prime step 100 and 101. Their next bridge is past the max bridge interval, between prime step 152 and 153.}
\end{figure}

To illustrate the operation of the Philos Trust Algorithm, we will analyze a sync list's bridge process over the period of prime step $k=98$ to $k=154$ for Figure~\ref{fig:normalexample} and Figure~\ref{fig:lateexample}. For Figure~\ref{fig:earlyexample}, we analyze a sync list's bridge process over the period of prime step $k=98$ to $k=113$. For the purposes of these examples, we will assume that the sync list performed a successful bridge at prime step $k=52$ and has completed every primary since that time. We offer three examples illustrating the sync list bridge process occurring
\begin{itemize}
    \item Perfect bridge process occurring at the max bridge interval (Figure~\ref{fig:normalexample}),
   \item Bridge process occurring too often combined with a missed primary sync (Figure~\ref{fig:earlyexample}), and
   \item Bridge process when a sync list bridges beyond the max bridge interval (Figure~\ref{fig:lateexample}).
\end{itemize}


    


\begin{figure}
    \centering
    \includegraphics[width=0.5\textwidth]{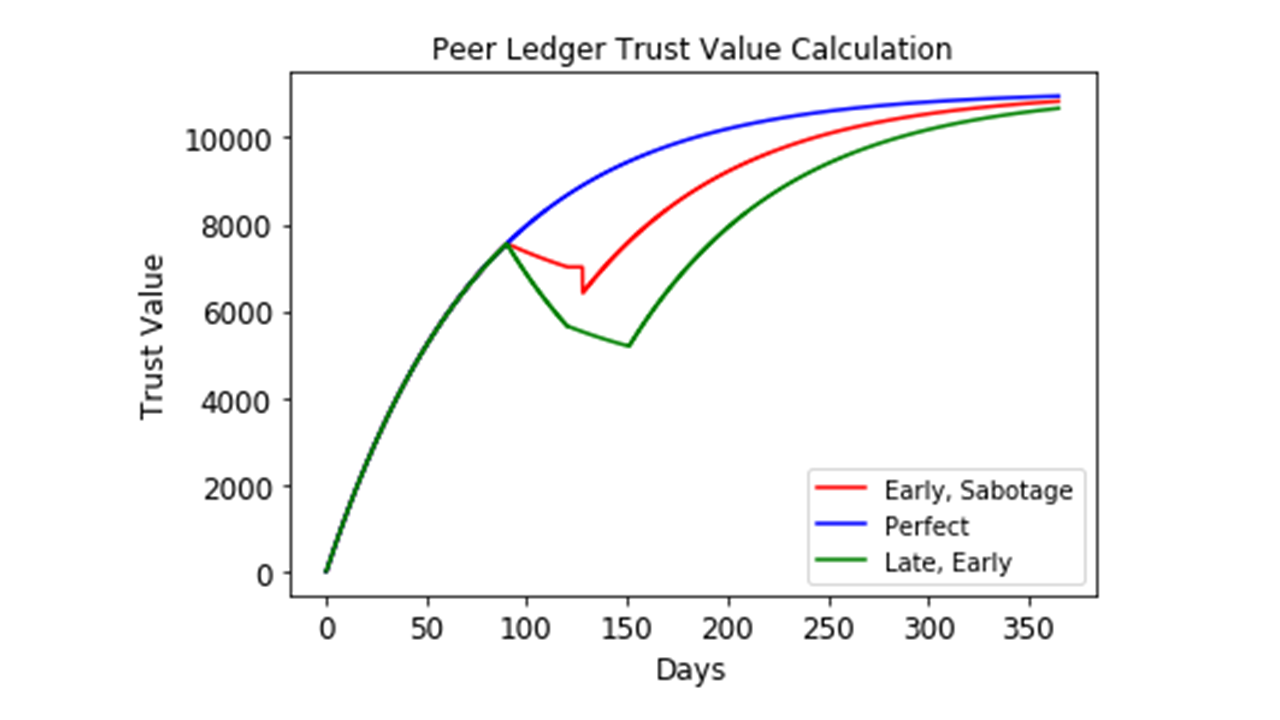}
    \caption{\footnotesize\label{fig:PLTrustPlot}Peer Trust Value Calculation as described in Section~\ref{ssec:examples}: An individual peer's trust value calculation over 365 days when the peer bridges (Blue) perfectly, (Red) perfectly, too early for 1 month, misses primary and is out for 1 week, perfectly from then on, (Green) perfectly, past max bridge interval for 1 month, too early for 1 month, perfectly rest of time. }
\end{figure}

Finally, Figure~\ref{fig:PLTrustPlot} illustrates the trust calculation for an individual peer over a period of 365 days considering three different scenarios of behavior. 
All parameters values are selected as in Table~\ref{table:parameters}.
The trust value depicted by the blue trace represents a peer that performs perfectly, bridging every 8 hours and never missing a primary. The trust value depicted in red represents a peer that performed perfectly for the first 3 months, consistently bridges too early for the next month (bridging every 4 hours and never missing a primary), misses a primary and is unmatched for one week and then performs perfectly for the remaining time. The trust value depicted by the green trace represents a peer that performs perfectly for the first 3 months, performs bridges after the max bridge interval for 1 month (bridging every 9 hours and never missing a primary), then performs bridges too early for the next month (bridging every 3 hours), and then performs perfectly for the remaining time.

First, note that any of these off-nominal behaviors caused significant and sudden reduction in the trust value, with the sharpest reduction occurring when primaries were missed (red), followed by bridging late (green around day 100), followed by bridging early (red around day 100).
Second, note that once the peers returned to perfect performance, the past failures are not ``forgotten'' immediately, but that within 6 months, all three peers have returned to nearly the same perfect trust value.

\section{Incentivizing Honest Behavior} \label{sec:incent}

\subsection{Sabotage by a Dishonest Peer} \label{subsec:sabotage}

Note that~\eqref{eq:trustvalue} shows that a peer's trust value is a function of both of its own behavior (how long since it last participated in a bridge) and the behavior of the other peers in its sync list. At any time, a peer can choose not to participate in a primary sync thereby preventing a bridge consensus block from occurring. Since the trust value of each peer in the sync list is only updated when bridging occurs, every peer's trust is affected if one peer causes a primary consensus interval to be missed.
The trust values of all the peers within a sync list fail to be updated when $S_M = 0$. However, the trust value algorithm still maintains the number of opportunities a peer has had to make primary syncs, $k-b_k^i$, called the primary sync counter.
This value is used as the exponential decay exponent which will cause a higher decay on a peer's previous trust value if a sync list has been sabotaged.
Therefore, a saboteur judges the effectiveness of their actions by calculating the advantage they would gain based on what the decay would have been with a bridge, without the successful primaries.

Peer $i$ gauges the effectiveness of sabotage by comparing ~\eqref{eq:trustvalue} to the value that would be recorded if trust were updated with $S_M = 0$. We term this the \emph{hypothetical trust,} given by 
\begin{equation} \label{eq:hypotheticalTrust}
    \bar{T}_i = T_i^-\beta^{k-b_k^i}.
\end{equation}

Analogously, we define the hypothetical fractional trust of peer $i$ as 
\begin{equation}
    \bar{\mathcal{T}}_i := \frac{\bar{T}_i}{\sum_{\ell\in N}\bar{T}_\ell}.
\end{equation}

To model the strategic behaviour of a peer that is considering sabotage, we define a utility function, where the peer can either behave honestly $H$ or can sabotage $S$. Formally, for $a\in\{{\rm H,S}\}$, the peer's utility function is:
\begin{equation}
    U_i(a) = \left\{
    \begin{array}{ll}
        {\mathcal{T}_i} & \mbox{ if }a={\rm H} \\
        \bar{\mathcal{T}}_i& \mbox{ if }a={\rm S}.
    \end{array}
    \right.
\end{equation}

Mathematically, we wish to show under what conditions would it be unprofitable for peer $i$ to sabotage their sync list. That is, when is $U_i(S)<U_i(H)$?





%

The main question to ask is under what circumstances is it better for a peer to intentionally cause the a primary sync to be missed, thereby sabotaging the sync list? This question thus reduces to when is the peer's hypothetical fractional trust greater than the peer's fractional trust ${\bar{\cal T}}_i>{\cal T}_i$.

These questions are answered by the theorem stated below and its corollaries.
%

\begin{theorem} \label{thm:main}
Let peer $i$ with fractional trust $\bar{\mathcal{T}}_i$ be a member of a sync list $M$ containing $|M|$ total ledgers. Peer $i$ is incentivized to be honest if and only if 
\begin{equation} \label{eq:theorem}
|M|\bar{\mathcal{T}}_i  \le 1.
\end{equation}

\end{theorem}

\noindent The proof of Theorem~\ref{thm:main} is presented in the Appendix.


Theorem~\ref{thm:main} states that the only way a peer can gain a dishonest advantage is if either it has a large trust value, or if it is in a large sync list.
Since trust values are intrinsically upper bounded by the equilibrium of~\eqref{eq:trustvalue} (see Theorem~\ref{theorem:equilibrium}), this allows us to calculate a universal upper bound on the size of a sync list which guarantees no peer can dishonestly game the system.
This upper bound is reported in the following corollary:

\begin{corollary} \label{corollary}
Let $M$ be a sync list, and let $L$ be the total raw trust in the system.
If 
\begin{equation} \label{eq:corollary}
    |M| \leq \frac{L\left(1-\beta^\Delta\right)}{\Delta},
\end{equation}
then every peer in $M$ is incentivized to be honest.
\end{corollary}

\noindent We present the proof of Corollary~\ref{corollary} in the Appendix; it exploits the fact that the raw trust value of every peer is upper-bounded by $T^*$ as provided by Theorem~\ref{theorem:equilibrium}.

Furthermore, it is possible to use a similar technique to derive an upper bound on the size of a sync list which is expressed explicitly in terms of the \emph{average performance of the system:}

\begin{corollary} \label{corollary2}
Let $M$ be a sync list, and let $T_{\rm ave}$ be the total average raw trust in the system. If we relate the average trust to the upper bound, then $|N|T_{\rm ave}=L$ and we can write Corollary~\ref{corollary} as
\begin{equation} \label{eq:corollary2}
    |M| \leq \frac{|N|T_{\rm ave}}{T^*}.
\end{equation}
\end{corollary}
\vspace{2mm}

The fraction $\frac{T_{ave}}{T^*}$ is the normalized average trust of the system and is a metric of the system performance. If this value is close to 1, then it can be surmised that all peers are performing well and we can fit almost all peers into one sync list and still not have any exploitation.

\section{Conclusions}

In this paper, we have presented a complete description of the new Philos Trust Algorithm, used to compute trust values for the Philos Marketplace blockchain system. We characterized the equilibrium the trust value calculation and showed that it corresponds exactly to perfect performance.
Furthermore, we analyzed a specific strategic exploit of the algorithm, and showed that it can be prevented simply by limiting the number of participants in a sync list.
Future work will focus on other possible exploitative behaviors, such as those by patient or malicious adversaries.

\bibliographystyle{ieeetr}
\bibliography{library,libraryPam}


\begin{appendix}[Theorem Proofs]

\subsubsection*{Proof of Theorem~\ref{theorem:equilibrium}}
We first prove that $T^*$ is an equilibrium; we then prove that trust never exceeds $T^*$; finally, we show that trust increases at every bridge and complete the proof of asymptotic convergence.

When peer $i$ performs perfectly (i.e. $S_M=\Delta$ and $k-b_k^i=\Delta$) then~\eqref{eq:trustvalue} becomes
\begin{equation} \label{eq:perfectTrust}
    T_i=T_i^-\beta^\Delta+\Delta.
\end{equation}
Now, let $T_i^-=T^*$ as defined in~\eqref{eq:equiltrust}.
Then substituting in~\eqref{eq:perfectTrust} we have
\begin{align}
    T_i &= \frac{\Delta\beta^\Delta}{1-\beta^\Delta}+\Delta \nonumber\\
        &= \Delta\left(\frac{\beta^\Delta}{1-\beta^\Delta} + 1\right) \nonumber\\
        &= \frac{\Delta}{1-\beta^\Delta} = T^*.
\end{align}
That is, $T^*$ is an equilibrium point of~\eqref{eq:perfectTrust}.

Now, note that we may use~\eqref{eq:equiltrust} to write $\Delta = T^*(1-\beta^\Delta)$, and then substitute into~\eqref{eq:perfectTrust} to obtain
\begin{align}
    T_i 
        &=T_i^-\beta^\Delta + T^*(1-\beta^\Delta) \nonumber \\
        &=T_i^-\beta^\Delta + T^*(1-\beta^\Delta) + T_i^-(1-\beta^\Delta) - T_i^-(1-\beta^\Delta) \nonumber \\
        &=T_i^- + (1-\beta^\Delta)(T^* - T_i^-). \label{eq:dynamics}
\end{align}
Whenever $T^*>T_i^-$, the term $(1-\beta^\Delta)(T^* - T_i^-)>0$, so~\eqref{eq:dynamics} shows two things: first, that $T_i>T_i^-$, proving ~\eqref{eq:trust increases}.
Second,~\eqref{eq:dynamics} shows that if a peer has perfect performance, then at each bridge its trust value moves a $(1-\beta^\Delta)$ fraction of the way towards $T^*.$
Since $(1-\beta^\Delta)<1$, this ensures that trust remains bounded: whenever $T_i^-\leq T^*$, it must be that $T_i\leq T^*$, proving~\eqref{eq:tstar bound}.

Finally, note that~\eqref{eq:trust increases},~\eqref{eq:tstar bound} together with the Monotone Convergence Theorem prove that trust asymptotically approaches its equilibrium value of $T^*$, completing the proof.
%
%
%
%
%
%
%
%
\hfill\QED

\vspace{5mm}

\subsubsection*{Proof of Theorem~\ref{thm:main}}
Formally, we must show that there is no benefit for an individual peer to disband or sabotage the sync list preventing the sync list from successfully completing primary syncs. 
We start with an expression that defines an individual peer's share of the total system trust or Fractional Trust.
We will denote this individual peer as peer number 1 contained within a sync list $M$ of $n$ peers. Additionally, we treat $T_j$ constant as they are peers not participating in the sync list. The fractional trust is given by
\begin{equation} \label{eq:Ledgertotaltrustshare}
\mathcal{T}_1=\frac{\bar{T}_1 + S_M}{(\bar{T}_1 + S_M) + \sum_{i=2}^{n}(\bar{T}_i + S_M) + \sum_{j\notin{M}}\bar{T}_j},
\end{equation}
where $S_M$ is the number of successful primaries made by the sync list since the last bridge and $j$ indexes all peers in the system other than those in $M$.
The fractional trust for each individual peer is based on the ratio of the individual peer's raw trust to the total trust of the system. 


For simplicity, we let $\bar{T}_M = \sum_{i=2}^{n}\bar{T}_i$ denote the remaining total sync list trust value, $\bar{T}_s = \sum_{j\notin{M}}\bar{T}_j$ denote the remaining total system trust, then~\eqref{eq:Ledgertotaltrustshare} becomes
\begin{equation} \label{eq:FractionalTrust}
\mathcal{T}_1=\frac{\bar{T}_1 + S_M}{\bar{T}_1 + |M|S_M + \bar{T}_M + \bar{T}_s}.
\end{equation}

Using~\eqref{eq:FractionalTrust}, we can write the normalized hypothetical trust simply by setting $S_M=0$:
\begin{equation} \label{eq:HypotheticalTrustNorm}
\bar{\cal T}_1=\frac{\bar{T}_1}{\bar{T}_1 + \bar{T}_M + \bar{T}_s}
\end{equation}


By definition, peer $1$ is incentivized to be honest if and only if ${\bar{\cal T}}_1\leq{\cal T}_1$.
Using~\eqref{eq:FractionalTrust} and~\eqref{eq:HypotheticalTrustNorm}, ${\bar{\cal T}}_1\leq{\cal T}_1$ can be written as
\begin{equation} \label{eq:TrustInequality}
\frac{\bar{T}_1}{\bar{T}_1 + \bar{T}_M + \bar{T}_s} \leq \frac{\bar{T}_1 + S_M}{\bar{T}_1 + |M|S_M + \bar{T}_M + \bar{T}_s}.
\end{equation}



Letting $L = \bar{T}_1 + \bar{T}_M + \bar{T}_s$ which denote the total system trust value, then the inequality can be written as
\begin{equation} \label{eq:inequality}
\frac{\bar{T}_1}{L} \leq \frac{\bar{T}_1 + S_M}{L + |M|S_M}.
\end{equation}


Whenever $S_M > 0$, ~\eqref{eq:inequality}  is equivalent to

\begin{equation} \label{eq:PLedgerTrust}
\bar{T}_1 \leq \frac{L}{|M|},
\end{equation}

showing that~\eqref{eq:theorem} holds if and only if peer 1 is incentivized to be honest. Note that since $\frac{\bar{T}_1}{L} = \bar{{\cal T}}_1$, the proof is obtained.
\hfill\QED

\vspace{5mm}

\subsubsection*{Proof of Corollary~\ref{corollary}}
It was shown in Theorem~\ref{theorem:equilibrium} that for every peer $i$, ${T}_i\leq\frac{\Delta}{1-\beta^\Delta}$.
Note that~\eqref{eq:trustvalue} and~\eqref{eq:hypotheticalTrust} imply that $\bar{T}_i\leq T_i,$ so that $\bar{T}_i\leq\frac{\Delta}{1-\beta^\Delta}$.
Given this, we have that
\begin{align}   \mathcal{\bar{T}}_i &= \frac{\bar{T}_i}{L}, \label{inequality1}\\
                &\le \frac{\Delta}{L\left(1-\beta^\Delta\right)}, \label{inequality2}\\
                        &\le \frac{1}{|M|} \label{inequality3},
\end{align}
where~\eqref{inequality1} is the definition of hypothetical fractional trust, ~\eqref{inequality2} is the upper bound on $\bar{T}_1$ and~\eqref{inequality3} is the assumption of Corollary \ref{corollary}. Inequalities~\eqref{inequality1}--\eqref{inequality3}  are equivalent to
\begin{equation}
    |M|\bar{\mathcal{T}}_i  \le 1.
\end{equation}
That is, Theorem~\ref{thm:main} guarantees that every peer  is incentivized to be honest and the proof is complete.
\hfill\QED

\end{appendix}

\end{document}